\documentclass{article}
\usepackage{graphicx,epsf}
\usepackage{amsmath}

\input{tcilatex}

\begin{document}

\noindent 
\vskip 0.3cm
%

\noindent 
\vskip 0.3cm
%

\noindent 
\vskip 0.3cm
%

\begin{center}
FRACTIONAL STATISTIC

\vskip 0.3cm
%

by

\vskip 0.3cm
%

M. C. Berg\`{e}re
\end{center}

\noindent 
\vskip 0.3cm
%

\noindent \thinspace \thinspace \thinspace \thinspace \thinspace \thinspace
\thinspace

\noindent%
%
\textbf{Abstract: }We improve Haldane's formula which gives the number of
configurations for $N$ particles on $d$ states in a fractional statistic
defined by the coupling $g=l/m$. Although nothing is changed in the
thermodynamic limit, the new formula makes sense for finite $N=pm+r$ with $p$
integer and $0<r\leq m.$ A geometrical interpretation of fractional
statistic is given in terms of ''composite particles''.

\noindent

\noindent 
\vskip 0.3cm
%

\noindent \textbf{I. Introduction}

\noindent \noindent 
\vskip 0.3cm
%

Fractional statistic was proposed by Haldane $\left[ 1\right] $ as a
generalization of Fermi--Dirac statistic where Pauli exclusion principle is
replaced by a more general exclusion principle: $g$ states are needed to add
one more particle to the system. Clearly, $g=0$ or $1$ corresponds
respectively to bosons or fermions statistics; however, this new statistic
is not restricted to integer values of $g$. For instance, if $g=1/m,$ the
exclusion principle tells that $m$ particles can be added to the system on
a single state. More generally, if $g=l/m$ the fractional statistic means
that $l$ states are needed to add $m$ particles to the system. Of course,
this interpretation makes sense for a large number of particles and states
(thermodynamic limit); however, a microscopic interpretation is difficult to
realize and for instance, we would like to understand what happens if we add
one more particle only to the system since in that case an additional
fractional number of states is meaningless. In $\left[ 2\right] $
Polychronakos tried to answer this question by arranging the states on a one
dimensional open lattice with the restriction that any two particles be at
least $g$ sites apart ($g$ integer); although this modifies the combinatoric
in the microscopic regime, it gives back Haldane's statistic in the
thermodynamic limit. Unfortunetely, his proposition together with the
assumption of factorizability of the partition function lead necessarily to
negative weights for some configurations (this fact was also observed before
by Nayak and Wilczek $\left[ 3\right] $ for $g=1/2)$. Later on, Chaturvedi
and Srinivasan $\left[ 4\right] $ showed that the fractional statistic was
not compatible with the factorizability of the partition function. For $%
g=1/2 $ and for an odd number of particles, they calculated the positive
fractional weight for any configuration. More recently Murthy and Shankar $%
\left[ 5\right] $ explained that all configurations were not a priori
allowed in the fractional statistic and they determine the constraints which
define the possible configurations. For sake of symmetrization over all
configurations these constraints can be forgotten at the price of
introducing positive fractional weights; they calculated these weights for $%
g=1/2$ $\,$with an odd number of particles (in agreement with the results of 
$\left[ 4\right] $) and for $g=1/3$ with a number of particles $N=3p+1.$

In this publication, we improve Haldane's formula which gives the number of
possible configurations for $N$ particles over $d$ states without changing
its thermodynamic limit; then, we give a geometrical interpretation of
fractional statistic for any number of particles and states. This
geometrical interpretation generalizes to all $g=l/m,$ to all number of
particles $N$ and to all number of states $d,$ the constraints of ref $\left[
5\right] $ over the allowed configurations. Finally, by symmetrization over the
configurations, we calculate the fractional weights in full generality.

\noindent 
\vskip 0.3cm
%

Given a set of states with energy $\epsilon _{i}$ and chemical potentiel $%
\mu ,$ we define the variables 
\begin{equation}
x_{i}=\exp \left( -\beta \left( \epsilon _{i}-\mu \right) \right)  \label{1}
\end{equation}
where $\beta $ is the inverse temperature 1/T. Then, the partition function
for the bosonic statistic is 
\begin{equation}
Z\left( x_{1},...,x_{d}\right) =\prod_{i=1}^{d}\frac{1}{1-x_{i}}%
=\sum_{\left\{ p_{i}\right\} }x_{1}^{p_{1}}...x_{d}^{p_{d}}  \label{2}
\end{equation}
where we sum over all integers $p_{i}\geq 0.$ If all energies are equal, the
partition function becomes 
\begin{equation}
Z(x)=\frac{1}{\left( 1-x\right) ^{d}}=\sum_{N=0}^{\infty }C_{d+N-1}^{N}x^{N}
\label{3}
\end{equation}
where $C_{d+N-1}^{N}$ is the number of configurations for $N$ particles and $%
d$ states. Similarly, the partition function for the fermionic statistic is

\begin{equation}
Z\left( x_{1},...,x_{d}\right) =\prod_{i=1}^{d}\left( 1+x_{i}\right)
=\sum_{\left\{ p_{i}=0,1\right\} }x_{1}^{p_{1}}...x_{d}^{p_{d}}  \label{4}
\end{equation}
and with equal energies we get

\begin{equation}
Z(x)=\left( 1+x\right) ^{d}=\sum_{N=0}^{\infty }C_{d}^{N}x^{N}  \label{5}
\end{equation}

Now, the partition function for \noindent Haldane's fractional statistic
when all energies are equal ($\epsilon _{i}=\epsilon ;x_{i}=x$) is defined
as 
\begin{equation}
Z\left( x\right) =\sum_{N=0}^{\infty }G_{d}^{N}\left( g\right) x^{N}
\label{6}
\end{equation}
where $g$ is a rational $l/m$ and where the number of configurations for $N$
particles on $d$ states is given by 
\begin{equation}
G_{d}^{N}\left( g\right) =C_{d+\left( 1-g\right) \left( N-1\right) }^{N}
\label{7}
\end{equation}

The above statistic interpoles between the bosonic statistic ($g=0$) and the
fermionic statistic ($g=1$). Of course, $G_{d}^{N}\left( g\right) $ must be
an integer number since it represents a number of configurations. However,
when $g=l/m$, the above formula has a meaning only if we restrict the number
of particles to be of the form 
\begin{equation}
N=pm+1  \label{8}
\end{equation}
(p integer) because of the factorial function included in the $C$ symbol (an
analytic continuation using the Euler $\Gamma $ function is wrong since it
gives a fractional number of configurations). Usually, this statistic is
understood in the large $N$ and large $d$ case (thermodynamic limit) where
we define the average number of particles per state as 
\begin{equation}
n=\lim d\rightarrow \infty \,\ \frac{\overline{N}}{d}  \label{9}
\end{equation}
where $\overline{N\text{ }}$is defined as the extremum over $N$ of $%
[G_{d}^{N}\left( g\right) x^{N}]$ (in this calculation, $\lg \left(
N!\right) \sim N\lg \left( \frac{N}{e}\right) $for large $N$). It is found
that 
\begin{equation}
n=\frac{1}{W\left( x,g\right) +g}<\frac{1}{g}  \label{10}
\end{equation}
where $W\left( x,g\right) $ is a positive quantity which satisfies the
equation $\left [ 6-9\right] $:

\begin{equation}
W^{g}\left( 1+W\right) ^{1-g}=\frac{1}{x}  \label{11}
\end{equation}
At small temperature ($T\rightarrow 0$) and when $\epsilon <\mu $ the
quantity $W\rightarrow 0$ so that the average number of particles per energy
level $n\rightarrow \frac{1}{g}$

\noindent 
\vskip 0.3cm
%

One purpose of this contribution is to generalize Haldane's fractional
statistic to all numbers of particles 
\begin{equation}
N=pm+r\,\,\,,\,\,0<r\leq m  \label{12}
\end{equation}
In the next section, we give a geometrical interpretation to the fractional
statistic for any $N$ and any $d$ and we prove that the corresponding number
of possible configurations for a given $g=l/m$ is given by the following
function 
\begin{equation}
F_{d}^{N}\left( l,m\right) =C_{d+N-1-lE\left( \frac{N-1}{m}\right) }^{N}
\label{13}
\end{equation}
where the function ''integer part'' is such that $E\left( \frac{N-1}{m}%
\right) =p.$

This fractional statistic gives the bosonic statistic for $l=0,$ the
fermionic statistic for $l=m=1$ and Haldane's fractional statistic for $%
N=pm+1.$ Of course, for $N$ large, the effect of the function ''integer part
''disappears and we get back the average number of particles per state $n$
as given by (10) and Ouvry's equation (11). It is interesting to note that
for $g=1,$ we obtain the fermionic statistic only if $l=m=1,$ but we obtain
different statistics if $l=m=k$ although they all coincide in the
thermodynamic limit.

The general idea for the geometrical interpretation is to take seriously for
any finite number of particles $N$, an organisation containing $p$
''composite particles'' (set of m particles and at least ($l-1)$ empty
states) plus one uncomplete ''composite particle'' containing $r$ particles and
any number of empty states. When we symmetrize this picture over all states,
we obtain a statistic of $(p+1)$ ''composite particles'' with fractional
weights for each configuration. Let us try to clarify this point:

In the bosonic statistics with states of different energy $\epsilon _{i}$,
the partition function may be written as (2) where we sum over all possible
monomials$\,x_{1}^{p_{1}}...x_{d}^{p_{d}}$ with weight 1 for all monomials.
In the fermionic statistic, the monomials are such that $p_{i}=0,1$ for all $%
i=1,...,d$ so that the weight for a given monomial is 1 or 0 accordingly. Of
course, for these two statistics, the partition function $Z\left(
x_{1},...,x_{d}\right) $ is symmetric in the variables $x_{i},$ and is
factorizable as 
\begin{equation}
Z\left( x_{1},...,x_{d}\right) =\prod_{i=1}^{d}Z\left( x_{i}\right)
\label{14}
\end{equation}

It is now known $\left[ 4,5\right] $ that for the fractional statistic, the
partition function is not factorizable. After S. Chaturvedi and V.
Srinivasan, we write 
\begin{equation}
Z\left( x_{1},...,x_{d}\right) =\sum_{\left\{ p_{i}\right\} }f\left(
p_{1},...,p_{d}\right) x_{1}^{p_{1}}...x_{d}^{p_{d}}  \label{15}
\end{equation}
where the weights $f\left( p_{1},...,p_{d}\right) $ are $\geq 0$ and
symmetric. Since we generate in (15) symmetric polynomials, we may introduce
the set of homogeneous symmetric polynomials 
\begin{equation}
M_{\lambda }\left( x\right) =\sum x_{1}^{\lambda _{1}}...x_{d}^{\lambda _{d}}
\label{16}
\end{equation}
attached to a given Young tableau $\lambda $ and where the sum runs over all
distinct permutations of $\left( \lambda _{1},...,\lambda _{d}\right) .$ The
surface $\left| \lambda \right| $ of the Young tableau represents the number 
$N$ of particles. Then 
\begin{equation}
Z\left( x_{1},...,x_{d}\right) =\sum_{\lambda }f_{\lambda }\,M_{\lambda
}\left( x\right)  \label{17}
\end{equation}

For bosonic statistic $f_{\lambda }=1$ for all $\lambda $ and for fermionic
statistic $f_{\lambda }=1$ for $\lambda =\left\{ 1^{N}\right\} $ and 0
otherwise. For fractional statistic, the weights $f_{\lambda }$ are
fractional.

These weights have been determined for $g=1/2$ and $N=2m+1$ by Chaturvedi
and Srinivasan; they were also obtained for $g=1/2,1/3$ and $N=2p+1,3p+1$
respectively by Murthy and Shankar who mentionned the existence of an
algorithm to calculate $f_{\lambda }$ for arbitrary $g=1/m$ and $N=pm+1.$

In section II, we give a geometrical interpretation of the fractional
statistic which leads to an intermediate non-symmetric partition function
and we calculate $F_{d}^{N}\left( g\right) $ for this construction. In
section III, we symmetrize the result of section II and calculate the
corresponding weights $f_{\lambda }.$

\noindent \textbf{II. Geometrical construction of the intermediate unsymmetric
partition function.}
\vskip 0.5cm
Let us give ourself $d$ states which can be drawn from the top first level
to the bottom $d^{th}$ level. Let us give ourself $N=pm+r$ $\,\,\left(
0<r\leq m\right) $ particles to be placed on the $d$ states in the following
way :

We define a ''composite particle'' as a set of $m$ particles and $\delta
\geq l$ states with the constraint that the $\left( l-1\right) $ bottom
states are empty and the $l^{th}$ state (from the bottom) has at least one
particle. We define an uncomplete ''composite particle'' as a set of $%
0<r\leq m$ particles with no constraint on the empty states. The situation
is described in fig.1 with $(p+1)$ ''composite particles'', the uncomplete
one being placed at the bottom. Clearly, the main constraint in this
construction is that a non empty state is entirely included inside a
''composite particle'' and cannot be splitted over several ''composite
particles'' (as showed in fig.2).

We now calculate the number of possible configurations for a given choice of 
$N,d,l,m.$ We first consider the uncomplete ''composite particle'' with $r$
particles on $\delta _{p+1}$ states. Since $r\leq m$ and since the empty
states have no constraint, the uncomplete ''composite particle'' satisfies a
bosonic statistic and the number of possible configurations is 
\begin{equation}
C_{\delta _{p+1}+r-1}^{r}  \label{18}
\end{equation}

We now consider a ''composite particle''; once the $\left( l-1\right) $
empty bottom states are fixed, we have a bosonic statistic for $m$ particles
on $\delta -\left( l-1\right) $ states but we have to subtract the number of
configurations where there is no particle on the $l^{th}$ state. The number
of possible configurations for a ''composite particle'' is 
\begin{equation}
C_{\delta -l+m}^{m}-C_{\delta -l+m-1}^{m}=C_{\delta -l+m-1}^{m-1}  \label{19}
\end{equation}

Consequently, given $p$ ''composite particles'' with $\delta _{i}$ states
satisfying 
\begin{equation}
\delta _{i}\geq l\,\,\,\,,\,i=1,...,p\,\,,  \label{20}
\end{equation}
and given one uncomplete ''composite particle'' with $\delta _{p+1}$ states
satisfying 
\begin{equation}
\delta _{p+1}\geq 1,  \label{21}
\end{equation}
such that the total number of states is
\begin{equation}
\sum_{i=1}^{p+1}\delta _{i}=d,  \label{22}
\end{equation}
then, the total number of possible configurations is 
\begin{equation}
F_{d}^{N}\left( l,m\right) =\sum_{\left\{ \delta _{i}\right\} }C_{\delta
_{p+1}+r-1}^{r}\prod_{i=1}^{p}C_{\delta _{i}-l+m-1}^{m-1}  \label{23}
\end{equation}
where we sum over all $\delta _{i}$'s defined above. To perform this sum, we
proceed as usual: we define the functional 
\begin{equation}
F\left( z_{1},...,z_{p+1}\right) =\sum_{\left\{ \delta _{i}\right\}
}C_{\delta _{p+1}+r-1}^{r}\,z_{p+1}^{\delta
_{p+1}-1}\,\prod_{i=1}^{p}C_{\delta _{i}-l+m-1}^{m-1}z_{i}^{\delta _{i}-l}
\label{24}
\end{equation}
where the sums over the variables $\delta _{i}$ satisfy (20,21) and run to $%
\infty .$ We get

\begin{equation}
F\left( z_{1,}...,z_{p+1}\right) =\frac{1}{\left( 1-z_{p+1}\right) ^{r+1}}%
\prod_{i=1}^{p}\frac{1}{\left( 1-z_{i}\right) ^{m}}  \label{25}
\end{equation}
If we choose all variables $z_{i}=z$, we get 
\begin{equation}
\sum_{d=pl+1}^{\infty }F_{d}^{N}\left( l,m\right) \,z^{d-pl-1}=\frac{1}{%
\left( 1-z\right) ^{N+1}}=\sum_{n=0}^{\infty }C_{N+n}^{N}\,z^{n}  \label{26}
\end{equation}
By identification in $z$ and because $p=E\left( \frac{N-1}{m}\right) $ we
proved that the above organization of the particles in ''composite
particles'' leads to a statistic defined by 
\begin{equation}
F_{d}^{N}\left( l,m\right) =C_{d+N-1-lE\left( \frac{N-1}{m}\right) }^{N}
\label{27}
\end{equation}
\vskip 0.3cm
We now construct the partition function for $N$ particles. To each state $i$
we associate a variable $x_{i}\,\,\,(i=1,...,d).$ Then, each allowed
configuration defines a monomial $x_{1}^{p_{1}}...x_{d}^{p_{d}}$ where $%
p_{i} $ is the occupation number of the state $i.$ The above geometrical
construction means that the partition function is 
\begin{equation}
Z_{N}\left( x_{1},...,x_{d}\right) =\sum_{\left\{ p_{i}\right\} \in \Lambda
}x_{1}^{p_{1}}...x_{d}^{p_{d}}  \label{28}
\end{equation}
The set $\Lambda $ of possible $p_{i}$'s is defined by three constraints:

1${{}^{\circ }})$%
\begin{equation}
\sum_{i=1}^{d}p_{i}=N  \label{29}
\end{equation}

2${{}^{\circ }})$ let 
\begin{equation}
\pi _{i}=\sum_{j\leq i}p_{j}\,\,\,\,\,\,\,\,i=1,...,d  \label{30}
\end{equation}
then, 
\begin{equation}
\left\{ m,2m,...,pm\right\} \subseteq \left\{ \pi _{i}\right\}  \label{31}
\end{equation}
where $N=pm+r\,\,\,\,\,\,(0<r\leq m).$

3${{}^{\circ }})$ we consider the indices $(i_{1},...,i_{p})$ such that $\pi
_{i_{q}}=qm,$ then 
\begin{equation}
p_{i_{q}+1}=p_{i_{q}+2}=...=p_{i_{q}+l-1}=0\,,\,\,\,\,\,\,\,\,q=1,...,p
\label{32}
\end{equation}
We proved above that the number of monomials defined by $\Lambda $ is $%
F_{d}^{N}(l,m).$
\vskip 0.3cm
The partition function in (28) is an intermediate partition function which
is non-symmetric in the variables $x_{1},...,x_{d}.$ A
non-symmetric partition function does not seem to contradict any
principle of thermodynamic; however, we may have to deal with
physical systems where symmetrization is needed without changing
the total number of configurations. Then, fractional weights is a
consequence of the symmetrization procedure described in next section.
Before closing this section, let us make two remarks:

1${{}^{\circ }})$ we are now in position to answer the question: how many
states do we need if we add one particle to the system? Let us consider a
system with $N=pm+1$ particles and $(l-1)p$ empty states.. Adding one more
particle does not necessarily change the number of states as it completes
partly the uncomplete ''composite particle'' (from $r=1$ to $r=2$). Adding $%
(m-1)$ particles does not necessarily change the number of states for the
same reason. Now, if $N=pm+m$, adding one more particle takes $l$ extra
states: $(l-1)$ empty states to form the new ''composite particle'' and one
more state to receive the added particle. Altogether, adding $m$ particles
to $N$ particles takes at least $l$ extra states; as a consequence,
the average occupation number per state is $n=N/d\,<m/l.$

2${{}^{\circ }})\,$in the thermodynamic limit when $N$ and $d\rightarrow
\infty $ and in the zero temperature limit $T\rightarrow 0,$ the average
occupation number per state $n\rightarrow 1/g=m/l.$ This situation occurs
only when all ''composite particles'' have $\left( l-1\right) $ empty states
and one state with $m$ particles.
\vskip 0.5cm
\noindent \textbf{III Fractional weights.}
\vskip 0.5cm
We wish to transform the partition function (28) into the symmetric form 
\begin{equation}
Z_{N}\left( x_{1},...,x_{d}\right) =\sum_{\left\{ \lambda ,\left| \lambda
\right| =N\right\} }f_{\lambda }\,\,M_{\lambda }\left( x_{1},...,x_{d}\right)
\label{33}
\end{equation}
where the symmetric polynomial $M_{\lambda }\left( x_{1},...,x_{d}\right) $
is defined in (16) and in such a way that the total number of configurations
for $N$ particles on $d$ states is unchanged 
\begin{equation}
Z_{N}\left( 1,...,1\right) =F_{d}^{N}\left( l,m\right)  \label{34}
\end{equation}

Clearly, the weights $f_{\lambda }$ are the ratio between the number of
monomials of $M_{\lambda }$ which belong to $\Lambda $ (defined in section
II) and the total number of monomials in $M_{\lambda }.$ This last number is
known to be 
\begin{equation}
M_{\lambda }\left( 1,...,1\right) =\frac{d!}{q_{0}!q_{1}!q_{2}!...q_{m}!}%
=C_{d}^{q_{0}}\left[ q_{1},...,q_{m}\right]  \label{35}
\end{equation}
where $q_{i}$ is the number of rows of length $i$ in $\lambda $ (that is the
number of states with i particles), and where the symbol $\left[
q_{1},...,q_{m}\right] $ is 
\begin{equation}
\left[ q_{1},...,q_{m}\right] =\frac{\left( \sum_{i=1}^{m}q_{i}\right) !}{%
\prod_{i=1}^{m}q_{i}!}  \label{36}
\end{equation}
The numbers $q_{i}$ satisfy two relations 
\begin{equation}
\sum_{i=0}^{m}q_{i}=d  \label{37}
\end{equation}

\begin{equation}
\sum_{i=1}^{m}i\,q_{i}=N  \label{38}
\end{equation}
\vskip 0.3cm
The problem of finding how many monomials of $M_{\lambda }$ are in $\Lambda $
is much more elaborate and is treated in the appendix. Let us simply
describe the result:

\begin{equation}
f_{\lambda }=\frac{\left[ t_{1},...,t_{K(m)}\right] \prod_{i=1}^{p}\Psi
_{\mu _{i}}\,\,\Phi _{\nu }\,\,\,\,f_{0}}{\left[ q_{1},...,q_{m}\right] }
\label{39}
\end{equation}
where $f_{0}$ is related to the statistic of the empty states 
\begin{equation}
f_{0}=\frac{C_{q_{0}}^{\left( l-1\right) p}}{C_{d}^{\left( l-1\right) p}}=%
\frac{C_{d-\left( l-1\right) p}^{q_{0}-\left( l-1\right) p}}{C_{d}^{q_{0}}}
\label{40}
\end{equation}
In (39), the integers $t_{i,\,\,}\Psi _{\mu _{i}}$ and $\Phi _{\nu }$ are
independant of the empty states so that we may define them in a system where$%
\,\,l=1$ and $\,q_{0}=0$. In that case, each ''composite particle'' $i$
defines a Young tableau $\mu _{i};$ for each Young tableau, we define the
set of integers $r_{1},...,r_{m}$ corresponding to the number of rows with
length $1,...,m$ (that is the number of states with $1,...,m$ particles) and
satisfying 
\begin{equation}
\sum_{i=1}^{m}ir_{i}=m  \label{41}
\end{equation}
Then, 
\begin{equation}
\Psi _{\mu _{i}}=\left[ r_{1},...,r_{m}\right]  \label{42}
\end{equation}
Similarly, the uncomplete ''composite particle'' defines a Young tableau $%
\nu $ such that the corresponding multiplicities $s_{1},...,s_{r}$ satisfy 
\begin{equation}
\sum_{i=1}^{r}is_{i}=r  \label{43}
\end{equation}
Then, 
\begin{equation}
\Phi _{\nu }=\left[ s_{1},...,s_{r}\right]  \label{44}
\end{equation}
Finally, we denote by $K(m)$ the number of Young tableaux of surface $m$ and
by $t_{1},...,t_{K(m)}$ the multiplicities in the chosen set $\left\{ \mu
_{1},...,\mu _{p,}(\nu \text{ if }r=m)\right\} .$ These multiplicities
satisfy 
\begin{equation}
\sum_{i=1}^{K(m)}t_{i}=E\left( \frac{N}{m}\right)  \label{45}
\end{equation}
Then, a combinatorial factor $\left[ t_{1},...,t_{K(m)}\right] $ is
generated in (39) when we symmetrise over the ''composite particles'' (and
eventually over the uncomplete one if $r=m$).
\vskip 0.3cm
We now calculate the weights $f_{\lambda }$ by application of the formula
(39) to four simple examples:

1${{}^{\circ }})\,\,$the case $N\leq m:$

in this case, $p=0$ so that $f_{0}=1.$ Also, we have $E\left( \frac{N}{m}%
\right) =0$ or $1$ so that $\left[ t_{1,}...,t_{K(m)}\right] =1.$ Finally, $%
s_{i}=q_{i}$ ($q_{r+1}=...=q_{m}=0$) so that $\left[ s_{1},...,s_{r}\right] $%
=$\left[ q_{1},...,q_{m}\right] .$ Consequently, all $f_{\lambda }=1$ and we
are in a bosonic situation.

2${{}^\circ})\,$the case $m=1:$

in this case, there exists only one possible Young tableau so that $\left[
r_{1}\right] =\left[ s_{1}\right] =\left[ q_{1}\right] =\left[ t_{1}\right]
=1.$ In that case, $\,f_{\lambda }=f_{0}$ depends only of the statistic of
the empty states.

3${{}^\circ})\,$the case $m=2:$

in this case, we have two Young tableaux satisfying $\left[ 0,1\right] =%
\left[ 2,0\right] =1;$ clearly, $\left[ s_{1},s_{2}\right] =1$ whether $r=1$
or $2.$ The total number of Young tableaux $\left[ 0,1\right] $ is given by $%
t_{2}=q_{2}$ so that $\left[ t_{1},t_{2}\right] =C_{E\left( \frac{N}{2}%
\right) }^{q_{2}}$. Consequently, 
\begin{equation}
f_{\lambda }=\frac{C_{E\left( \frac{N}{2}\right) }^{q_{2}}\,f_{0}}{%
C_{q_{1}+q_{2}}^{q_{2}}}  \label{46}
\end{equation}
If $l=r=1,$ we obtain the results of ref.$\left[ 4,5\right] .$

4${{}^\circ})\,$the case $m=3$:

in this case, we have three Young tableaux satisfying $\left[ 0,0,1\right] =%
\left[ 3,0,0\right] =1$ and $\left[ 1,1,0\right] =2.$ On the other hand, $%
\left[ s_{1},s_{2},s_{3}\right] =1$ for $r=1$ or $2.$ The total number of
tableaux\thinspace $\left[ 0,0,1\right] $ is $t_{3}=q_{3},$ the total number
of tableaux $\left[ 1,1,0\right] $ is $t_{2}=q_{2}$ if $r=1$ or $3$, while
it is $t_{2}=q_{2}-1$ if $r=2$ and $\left[ s_{1},s_{2}\right] =\left[ 0,1%
\right] $. Consequently, 
\begin{equation}
f_{\lambda }=\frac{\left[ E\left( \frac{N}{3}\right) -q_{2}-q_{3}+\eta
,q_{2}-\eta ,q_{3}\right] \,2^{q_{2}-\eta }\,f_{0}}{\left[ q_{1},q_{2},q_{3}%
\right] }  \label{47}
\end{equation}
where $\eta =1$ if $r=2$ and $\left[ s_{1},s_{2}\right] =\left[ 0,1\right] $
and $0$ otherwise. Again, for $l=r=1,$ we obtain the result of ref.$\left[ 5%
\right] .$
\vskip 0.5cm
\noindent \textbf{Acknowledgments}
\vskip 0.5cm
I wish to thank E. Guitter for his organization of the combinatoric and D.
Bernard for several useful remarks and for a careful reading of
the manuscript.
\vskip 0.5cm
\noindent \textbf{Appendix}
\vskip 0.5cm
We now calculate the number of monomials of $M_{\lambda }\left(
x_{1},...,x_{d}\right) $ which are generated by the geometrical construction
of section II, that is the number of monomials which belong to $\Lambda $
defined in (29-32)$.$

The generating functional for at most $m$ particles on one state can be
written as 
\begin{equation}
F\left( \alpha ,x\right) =\alpha _{0}+\alpha _{1}x+...+\alpha _{m}x^{m}
\label{48}
\end{equation}

More generally, the generating functional for at most $m\,$particles per
state over $\delta $ states is 
\begin{equation}
\prod_{i=1}^{\delta }F\left( \alpha ,x_{i}\right) =\sum_{\lambda \in \left[
\delta *m\right] }\alpha _{\lambda _{1}}...\alpha _{\lambda _{\delta
}}\,M_{\lambda }\left( x_{1},...,x_{\delta }\right)  \label{49}
\end{equation}
where we sum over all Young tableaux $\lambda $ inside the rectangle $\left[
\delta *m\right] .$

At this stage, we find useful to introduce the multiplicities $s_{i}$ which
is the number of rows of length $i$ in $\lambda $; these multiplicities
satisfy 
\begin{equation}
\sum_{i=0}^{m}s_{i}=\delta  \label{50}
\end{equation}

\begin{equation}
\sum_{i=1}^{m}i\,s_{i}=\left| \lambda \right|  \label{51}
\end{equation}
Then 
\begin{equation}
\prod_{i=1}^{\delta }F\left( \alpha ,x_{i}\right) =\sum_{\lambda \in \left[
\delta *m\right] }\prod_{i=0}^{m}\alpha _{i}^{s_{i}}\,M_{\lambda }\left(
x_{1},...,x_{\delta }\right)  \label{52}
\end{equation}
The generating functional which gives the number of monomials is 
\begin{equation}
\left[ F\left( a,x\right) \right] ^{\delta }=\sum_{\lambda \in \left[ \delta
*m\right] }\left[ s_{0},...,s_{m}\right] \,\prod_{i=0}^{m}\alpha
_{i}^{s_{i}}\,x^{\left| \lambda \right| }  \label{53}
\end{equation}
where the symbol $\left[ s_{0},...,s_{m}\right] =M_{\lambda }\left(
1^{\delta }\right) \,$is given in (36).
\vskip 0.3cm
In the following, we introduce a generating functional for each ''composite
particle'' and also for the uncomplete ''composite particle''. To simplify
the writing, we define the generating functional $\Phi \left( \alpha
,x\right) $ which takes into account any number of states 
\begin{equation}
\Phi \left( \alpha ,x\right) =\sum_{\delta =0}^{\infty }\,\left[ F\left(
\alpha ,x\right) \right] ^{\delta }\,y^{\delta }=\frac{1}{1-F\left( \alpha
,x\right) \,y}  \label{54}
\end{equation}
and we find convenient to expand it under the form 
\begin{equation}
\Phi \left( \alpha ,x\right) =\sum_{n=0}^{\infty }\frac{\left( \alpha
_{1}x+...+\alpha _{m}x^{m}\right) ^{n}y^{n}}{\left( 1-\alpha _{0}y\right)
^{n+1}}  \label{55}
\end{equation}
so that the empty states get separated from the others. In order to describe
the $r$ particles of the uncomplete ''composite particle'', we must collect
in (55) all terms in $x^{r}.$ We get 
\begin{equation}
\Phi _{r}\left( \alpha \right) =\sum_{\left\{ \nu ,\left| \nu \right|
=r\right\} }\left[ s_{1},...,s_{r}\right] \,\prod_{i=1}^{m}\alpha
_{i}^{s_{i}}\,\frac{y^{n\left( \nu \right) }}{\left( 1-\alpha _{0}y\right)
^{n\left( \nu \right) +1}}  \label{56}
\end{equation}
where we sum over the Young tableaux $\nu $ with multiplicities $s_{i}$
satisfying 
\begin{equation}
\sum_{i=1}^{r}is_{i}=r  \label{57}
\end{equation}
\begin{equation}
\sum_{i=1}^{r}s_{i}=n\left( \nu \right)  \label{58}
\end{equation}
An expansion in $y$ of (56) gives for the term in $y^{\delta }$ the same
result as (53) restricted to $\left| \lambda \right| =r.\,$\thinspace We now
describe the ''composite particle'' from the generating functional 
\begin{equation}
\alpha _{0}^{l-1}\,(\alpha _{1}x+...+\alpha _{m}x^{m})\,\left[ F\left(
\alpha ,x\right) \right] ^{\delta -l}\,y^{\delta }  \label{59}
\end{equation}
which takes into account the specific structure of the ''composite
particle''. After summation over $\delta $ from $l$ to $\infty ,$ we obtain
the corresponding functional 
\begin{equation}
\Psi \left( \alpha ,x\right) =\frac{\alpha _{0}^{l-1}\left( \alpha
_{1}x+...+\alpha _{m}x^{m}\right) \,y^{l}}{1-F\left( \alpha ,x\right) \,y}
\label{60}
\end{equation}
which may also be expanded as 
\begin{equation}
\Psi \left( \alpha ,x\right) =\sum_{n=0}^{\infty }\frac{\alpha
_{0}^{l-1}\left( \alpha _{1}x+...+\alpha _{m}x^{m}\right) ^{n+1}\,y^{n+l}}{%
\left( 1-\alpha _{0}y\right) ^{n+1}}  \label{61}
\end{equation}
Consequently, the ''composite particle'' with $m$ particles has for
generating functional 
\begin{equation}
\Psi _{m}(\alpha )=\alpha _{0}^{l-1}\sum_{\left\{ \mu ,\left| \mu \right|
=m\right\} }\left[ r_{1},...,r_{m}\right] \prod_{i=1}^{m}\alpha
_{i}^{r_{i}}\,\frac{y^{n\left( \mu \right) +l-1}}{\left( 1-\alpha
_{0}y\right) ^{n\left( \mu \right) }}  \label{62}
\end{equation}
where we sum over the Young tableaux $\mu $ with multiplicities $r_{i}$
satisfying 
\begin{equation}
\sum_{i=1}^{m}ir_{i}=m  \label{63}
\end{equation}
\begin{equation}
\sum_{i=1}^{m}r_{i}=n\left( \mu \right)  \label{64}
\end{equation}

The generating functional for $p$ ''composite particles'' and one uncomplete
''composite particle'' with $r$ particles is given by 
\begin{equation}
\Psi _{m}^{p}(\alpha )\,\Phi _{r}(\alpha )=\alpha _{0}^{\left( l-1\right)
p}\sum_{\left\{ \mu _{1},..,\mu _{p},\nu \right\} }\Phi _{\nu
}\prod_{i=1}^{p}\Psi _{\mu _{i}}\,\prod_{i=1}^{m}\alpha _{i}^{q_{i}}\,\frac{%
y^{d-q_{0}+p(l-1)}}{\left( 1-\alpha _{0}y\right) ^{d-q_{0}+1}}  \label{65}
\end{equation}
where $\Psi _{\mu _{i}}=\left[ r_{1},...,r_{m}\right] $ for the Young
tableau $\mu _{i},\,\,\Phi _{\nu }=\left[ s_{1},...,s_{r}\right] $ for the
Young tableau $\nu $ and where the $q_{i}$'s are defined in (37,38). If we
develop (65) in powers of $y$ to get the term in $y^{d}$ where $d$ is the
total number of states, we obtain for the total number of monomials
corresponding to the chosen set of Young tableaux $\left\{ \mu _{1},...,\mu
_{p},\nu \right\} $ 
\begin{equation}
C_{d-\left( l-1\right) p}^{q_{0}-\left( l-1\right) p}\,\sum_{\left\{ \mu
_{1},...,\mu _{p},\nu \right\} }\Phi _{\nu }\,\prod_{i=1}^{p}\Psi \mu _{i}
\label{66}
\end{equation}
Finally, any permutation of the tableaux $\mu _{i}$ (and eventually $\nu $
if $r=m$) contributes as well to the monomials of $M_{\lambda }\left(
x_{1},...,x_{d}\right) $ which belong to $\Lambda ;$ these permutations
generate a combinatoric factor equal to $\left[ t_{1},...,t_{K(m)}\right] $
where $K(m)$ is the number of Young Tableaux with surface $m$ and the
integers $t_{i}$ are their multiplicities (satisfying (45)) in the chosen
set $\left\{ \mu _{1},...,\mu _{p},(\nu \text{ if }r=m)\right\} $. This ends
the calculation of $\,f_{\lambda }$ as given in (39).$\,$
\vskip 0.5cm
$\left[ 1\right] $ F.D.M. Haldane, Phys. Rev. Lett. \textbf{67}, 937 (1991).

$\left[ 2\right] $ A. P. Polychronakos, Phys. Lett. \textbf{B365}, 202
(1996).

$\left[ 3\right] $ C. Nayak and F. Wilczek, Phys. Rev. Lett. \textbf{73},
2740 (1994).

$\left[ 4\right] $ S. Chaturvedi and V. Srinivasan, Phys. Rev. Lett. \textbf{%
78}, 4316 (1997).

$\left[ 5\right] $ M. V. N. Murthy and R. Shankar, IMSc/99/01/02,
Cond-mat/9903278.

$\left[ 6\right] $ B. Sutherland, J. Math. Phys. {\bf 12}, 251
(1971).

$\left[ 7\right] $ Y.S. Wu, Phys. Rev. Lett. {\bf 73}, 922
(1994).

$\left[ 8\right] $ D. Bernard, Les Houches, Session LXII (1994).

$\left[ 9\right] $ A. Dasni\`ere de Veigy and S. Ouvry, Phys.
Rev. Lett. {\bf 72}, 600 (1994).

\begin{figure}
\centerline{\includegraphics[width=15cm]{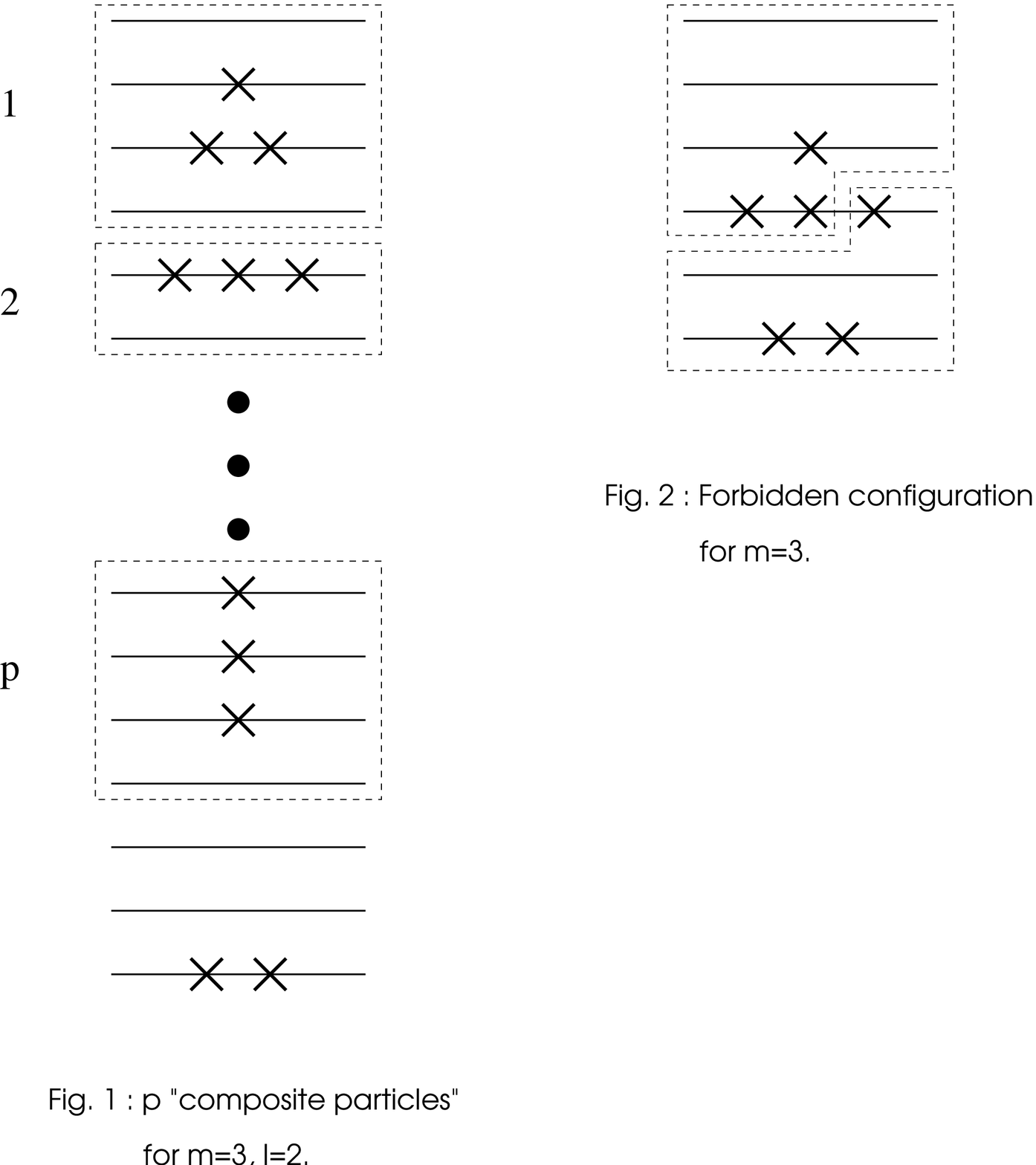}}
\end{figure}
\end{document}